\documentclass[twocolumn,aps,prb,groupedaddress]{revtex4-2}

\usepackage{chngcntr}

\usepackage{graphicx}
\usepackage{color}
\usepackage{amsmath}
\usepackage{enumitem}
\usepackage{amssymb}
\usepackage{hyperref}
\usepackage{cancel}
\usepackage{ulem}
\usepackage{multirow}

\newcommand{\be}{\begin{equation}}
	\newcommand{\ee}{\end{equation}}

\newcommand{\bea}{\begin{eqnarray}}
	\newcommand{\eea}{\end{eqnarray}}

\newcommand{\p}{\partial}

\newcommand{\la}{\left\langle}
\newcommand{\ra}{\right\rangle}

\renewcommand{\vec}[1]{{\boldsymbol #1}}

\begin{document}

\title{Two-dimensional electron gases as non-Newtonian 
fluids}
%
\author{Serhii Kryhin and Leonid Levitov}
\affiliation{Department of Physics, Massachusetts Institute of Technology, Cambridge, MA 02139}

\begin{abstract}
Two-dimensional electron systems offer an appealing platform to explore long-lived excitations arising due to collinear carrier scattering enabled by 
phase-space constraints at the Fermi surface. 
Recently it was found that these effects can boost excitation lifetimes over the fundamental bound set by Landau's Fermi-liquid theory by a factor as large as $(T_F/T)^\alpha$ with $\alpha\approx 2$.
Long-lived degrees of freedom possess the capability to amplify the response to weak perturbations, producing lasting collective memory effects. This leads to 
non-Newtonian hydrodynamics in 2D electron fluids driven by multiple 
viscous modes with scale-dependent viscosity. 
We describe these modes 
as Fermi surface modulations of odd parity evolving in space and time, and discuss their implications for experimental studies of electron hydrodynamics.
\end{abstract}

\date{\today}

\maketitle

Recent years have seen a surge of interest in Gurzhi's electron hydrodynamics \cite{Gurzhi1968} 
as a framework to describe 
transport in quantum materials at diverse length and time scales 
\cite{Guerrero-Becerra2019,Hasdeo2021,Muller2009,Principi2016,Scaffidi2017, Narozhny2019,Alekseev2020,Toshio2020,Narozhny2021,Tomadin2014, Principi2016,Lucas2018,Qi2021,Cook2021,Valentinis2021a,Valentinis2021b, DasSarma2022,
HGuo2017,AShytov2018,Nazaryan2021}. 
However, despite this burgeoning interest, the fundamental question of how an orderly hydrodynamic behavior on macroscales stems from a chaotic dynamics due to interactions and collisions on microscales, in particular the role of the quantum effects, has received relatively little attention. The situation is well understood for classical gases, where all moments of momentum distribution not explicitly protected by conservation laws are extremely fragile, being quickly erased by particle dynamics after just a few ($N\sim 1$) collisions [see, e.g., \cite{Liboff2003}]. To the contrary, as discussed below, quantum gases and liquids can 
feature surprising collective memory effects occurring over the span of $N\gg 1$ successive collisions with $N$ rapidly diverging at low temperatures.  The long-time 
dynamics in such systems cannot be captured by a conventional hydrodynamic description that relies on a closed set of equations for classically-conserved quantities such as local flow velocity, particle density and temperature. Instead, a full description must account for memory effects due to 
nonclassical quantities that are not protected by microscopic conservation laws but nonetheless feature abnormally long lifetimes. This behavior bears a resemblance to that observed in non-Newtonian fluids, in which collective microscopic memory effects lead to viscosity that varies with scale\cite{non-newtonian_fluids}.

 \begin{figure}[t]
\centering
\includegraphics[width=0.95\columnwidth]{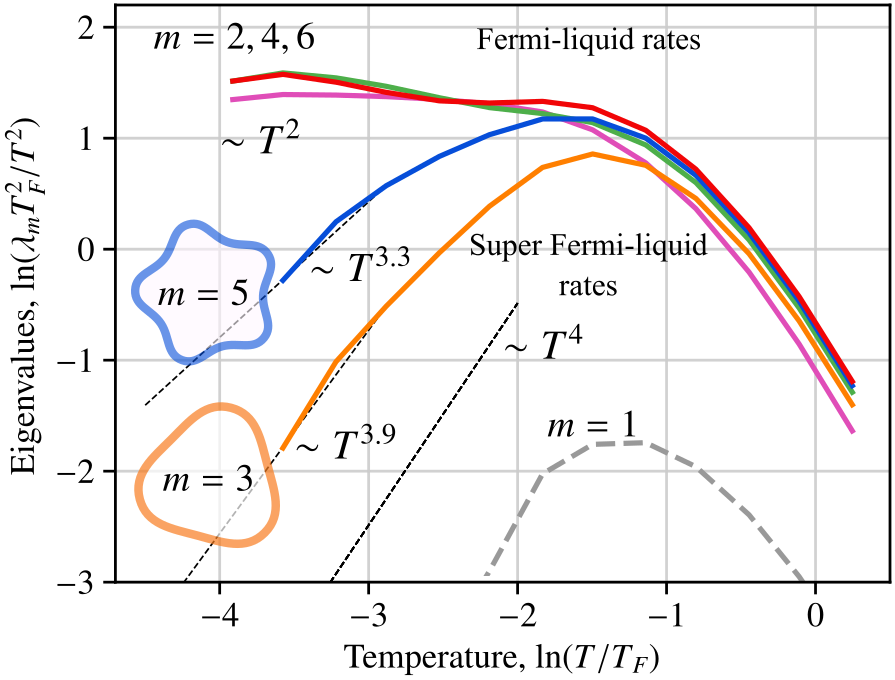} 
\caption{Decay rates for different angular harmonics of 
electron velocity distribution, scaled by $T^2$, vs. temperature (adapted from \cite{Kryhin2022}). 
Shown are dimensionless eigenvalues $\lambda_m$ related to the decay rates through $\gamma_m=A p_F^2 \lambda_m $, see Eq.(32) in \cite{Kryhin2022}.
Double-log scale is used to facilitate comparison of 
disparate time scales. 
Decay rates for even-$m$ harmonics obey a $T^2$ scaling at $T\ll T_F$. 
Decay rates for odd-$m$ harmonics are suppressed below the even-$m$ rates,  
showing ``super-Fermi-liquid'' scaling strongly deviating from $T^2$. Odd-$m$ decay rates can be approximated as $T^\alpha$ with $\alpha>2$. A small nonzero rate found for velocity mode $m=1$ is a parasitic effect (see text). 
An even/odd asymmetry in the rates and the suppression of decays for odd-$m$ harmonics are clearly seen below $T \approx 0.2 T_F$. 
}
\vspace{-5mm}
\label{fig:fig1}
\end{figure}

An early indication that nonclassical long-lived quantities 
may exist in electron systems was provided by another important paper by Gurzhi \cite{Gurzhi1995} (see also \cite{Buhmann2002}). This work emphasized the role of collinear scattering due to head-on collisions in 2D electron systems.
Curiously, it was thought at the time that collinear scattering shortens quasiparicle lifetimes through speeding up the excitations' decay by a non-Fermi-liquid log factor 
\cite{Hodges1971,Chaplik1971,Bloom1975,Giuliani1982,
Zheng1996,Menashe1996,Chubukov2003}. 
Subsequent work cleared this misconception and linked the collinear scattering in 2D systems to long-lived excitations, opening way to explore a variety of collective memory effects \cite{Ledwith2017,tomogrph,Ledwith2019}.
These results came at a crucial juncture as 2D systems have become the focal point of ongoing efforts to achieve electron hydrodynamics  \cite{Sulpizio2019,Ku2020,Braem2018,Vool2021,Aharon-Steinberg2022}. From theoretical standpoint, the properties of 2D systems were found to lie somewhere between those of 3D and 1D systems, being sharply distinct from both. 
For 3D systems, the Fermi-liquid theory confirms Boltzmann's short-time memory picture with the onset of hydrodynamics occurring after $\sim 1$ quasiparticle collisions \cite{BaymPethick}. 
The 1D systems feature manifestly non-Boltzmann behavior, described by the Luttinger-liquid theory that predicts integrable non-ergodic behavior that extends to arbitrarily large times and distances \cite{Giamarchi_book}. 
The unique behavior in 2D Fermi systems, which is 
due to the dominant role of head-on collisions 
\cite{Gurzhi1995,Buhmann2002,Ledwith2017}, deviates strongly from that in both 3D and 1D systems. 


Additionally, it is worth noting that the unique collinear nature of collisions at a 2D Fermi surface, leading to electron/hole backscattering with significant angular memory, has been referred to as 'tomographic' dynamics in previous studies (Refs. \cite{Ledwith2017, tomogrph, Ledwith2019}). However, this paper will not address these specific aspects of the problem. Accordingly, to avoid any potential confusion, we will refrain from using the term 'tomographic dynamics' throughout this paper.

The nonclassical long-lived excitations underpinning this behavior 
are odd-$m$ harmonics of the velocity distribution of fermions perturbed away from equilibrium. 
Recent microscopic analysis of quasiparticle scattering at a circular Fermi surface (FS) predicts quenching of the Landau $T^2$ damping for such modes \cite{Kryhin2022}. As illustrated in Fig.\ref{fig:fig1}, the quenching 
sets in fairly abruptly as temperature is lowered. Low-lying excitations in this system can be viewed as FS modulations evolving in space and time as
\begin{align}
\delta f(\vec p,\vec r,t)\sim \sum_{m\,{\rm odd}} 
\alpha_m\cos m\theta +\beta_m \sin m\theta
,
\end{align}
where $\theta$ is the angle parameterizing the FS and, for conciseness, we suppressed the dependence of the coefficients $\alpha_m$ and $\beta_m$ on position, time and particle energy. The microscopic decay rates pictured in Fig.\ref{fig:fig1} govern dynamics of spatially-uniform excitations, $\alpha_m$, $\beta_m\sim e^{-\gamma_m t}$. 
At low temperatures $T\ll T_F$ the lifetimes of odd-$m$ modes greatly exceed those for the even-$m$ ones and show strong departure from the conventional $T^2$ scaling. 

The decay rates in Ref.\cite{Kryhin2022} were obtained by a direct calculation that treats quasiparticle scattering exactly, using a method that does not rely on the small parameter $T/T_F\ll1$. 
The good accuracy of the method used in \cite{Kryhin2022} can be benchmarked by a residual $\gamma_m$ value for the velocity mode, $m=1$, that must vanish due to momentum conservation in two-body collisions. Here it takes a small nonzero value due to the effects of sampling the $p$ space and inaccuracies in calculating overlaps between different states. 
The odd-$m$ decay rates found for $m>1$ display scaling $\gamma_m\sim T^\alpha$ with super-Fermi-liquid exponents $\alpha>2$. The exponent $\alpha$ values were found to be 
close to $4$, indicating a strong suppression of the odd-$m$ rates 
compared to the even-$m$ rates, $\gamma_{\rm odd}/\gamma_{\rm even}\sim (T/T_F)^2$. This defines a new hierarchy of lifetimes for collective modes, leading to hydrodynamics with non-Newtonian (scale-dependent) viscosity. 

\section{
Odd-parity Fermi surface modulations}

To illustrate these properties we first focus on 
the two odd-$m$ harmonics with longest lifetimes, $m=\pm 1$ and $\pm 3$,  leaving out for the time being the harmonics 
with higher $m$. These two harmonics are singled out by 
a peculiar dependence of the lifetimes of excitations with different angular momenta $m$ on temperature and $m$ parity and magnitude that renders some excitations more long-lived than others. 
One characteristic property which makes odd-$m$ excitations long-lived \cite{Ledwith2019} is the even/odd difference in temperature scaling ($T^2$ for even $m$ and $T^4$ for odd  $m$). 
Another key property is that the odd-$m$ excitations with not too large $m$ display a steep $m^4$ dependence on $m$. This $T$ and $m$ dependence can be summarized as:
\begin{equation}\label{eq:OddAndEvenRates}
	 \gamma_{m\,{\rm odd}} \sim g \frac{T^4}{T_F^4} m^4 \ln m,\quad \gamma_{m\,{\rm even}} \sim g \frac{T^2}{T_F^2} \ln m,
\end{equation}
where $g$ is a dimensional factor that depends on the band structure \cite{Kryhin2022}. The two branches of excitations, even-$m$ and odd-$m$, are clearly set apart at not too large $m$. However, as $m$ grows, the even-$m$ and odd-$m$ rates eventually merge into a parity-blind mode at a high $m\gtrsim m_*\approx\sqrt{T_F/T}$. This behavior can be seen clearly in Fig. \ref{fig:fig1}. Naturally, the dependence in Eq.\eqref{eq:OddAndEvenRates} only applies to the part of momentum distribution that decays due to electron-electron (ee) collisions. This encompasses all harmonics with the exception of those with $m=0,\pm1$ that remain unchanged under ee collisions owing to the conservation of particle number and momentum. 


As a result, the velocity mode $m=\pm 1$ and the third-harmonic mode $m=\pm 3$ emerge as the longest-lived excitations. The former is damped only by momentum-relaxing scattering by disorder or phonons but not by ee collisions, the latter is damped by ee collisions more weakly than any higher-$m$ excitation (see Fig.\ref{fig:fig1}). 
%
In this two-mode approximation, the shear modes relevant for hydrodynamics take the form 
\be\label{eq:two_modes}
\delta f(\vec p,\vec r,t)\sim \beta_1(\vec r,t)\sin\theta+\beta_3(\vec r,t)\sin3\theta
, 
\ee
with 
$\theta$ measured from $\vec k$ direction, with a harmonic dependence $\beta_m(\vec r,t)\sim e^{i\vec k\vec r-i\omega t}$.  Here $\vec k$ is the wavenumber describing the mode spatial dependence. 
After integrating out the fast-relaxing modes with $m\ne 1,3$, we will arrive at 
two different hydrodynamic regimes that arise due to the appearance of a new timescale set by $\gamma_3$: 
\begin{enumerate}[label=(\roman*)]
\item the short-time regime $\omega, \, \nu k^2\gg\gamma_3$, and 
\item the long-time regime $\omega, \, \nu k^2\ll\gamma_3$,
\end{enumerate}
where  $\nu$ is the kinematic viscosity, $\nu=v_F^2/4\gamma_2$. 
At relatively short times $t\ll 1/\gamma_3$, i.e., in the first regime, the modes in Eq.\,\eqref{eq:two_modes} coupled by finite-$k$ effects discussed below yield two distinct viscous modes. These modes, found by diagonalizing a $2\times 2$ mode coupling problem, feature universal viscosity values
\be\label{eq:nu1,nu2}
\nu_1=\nu\frac{3+ \sqrt{5}}2,\quad
\nu_2=\nu\frac{3- \sqrt{5}}2
, 
\ee
satisfying $\nu_1>\nu>\nu_2$ and having 
a large ratio $\nu_1/\nu_2=(3+\sqrt{5})/(3-\sqrt{5})\approx 6.9$. 

The qualitative picture underpinning this regime is that 
quasiparticles collide at a normal Landau $T^2$ rate that can be estimated as $\gamma_2$, which leads to their diffusion in space with the diffusivity $\nu$. This diffusion, however, rather than being memory-erasing is of a memory-preserving character. 
Indeed, while the even-$m$ part of velocity distribution relaxes at relatively short times $\sim \gamma_2^{-1}$, the odd-$m$ part remains unrelaxed, producing memory effects manifested in a new hydrodynamic mode. 

In the second regime the conventional viscous (Newtonian) hydrodynamics is restored. This happens at distances such that $\nu k^2$ is smaller than $\gamma_3$:  
\be
L>\frac{v_F}{\sqrt{2\gamma_2\gamma_3}}
.
\ee 
These lengthscales are reached by a particle after undergoing many collisions, with the typical collision numbers estimated as
\be
N= 
\sqrt{\frac{\gamma_2}{2\gamma_3}} \sim \frac{T_F}{T} \gg 1
.
\ee
The number of successive collisions required for the memory of a microstate to be erased diverges in the low-$T$ limit, 
indicating a sharply non-Boltzmann behavior manifested as non-Newtonian hydrodynamics with scale-dependent viscosity. 


It is important to note that the inequality $N\gg1$ and the divergence of $N$ as $T$ approaches 0 should not be interpreted as evidence that electron-electron collisions are completely ineffective at erasing microstate memory. In contrast, the even-parity component of the momentum distribution follows the conventional Boltzmann collision behavior and becomes negligible after only around $N\sim 1$ collisions. On the other hand, the odd-parity component of the momentum distribution exhibits nonclassical behavior. Due to the phase space limitations inherent in collinear scattering, it requires an unusually high number of successive collisions to relax.

Turning to the analysis, we consider the most general framework describing electrons' velocity distribution by a kinetic equation linearized about equilibrium, 
\begin{equation}\label{eq:BKEgeneral}
	\left( \partial_t + \vec v\cdot\nabla \right) \delta f - I[\delta f] = - e \vec E(r) \cdot \frac{\p f_0}{\p \vec p}
	.
\end{equation}
Here $I$ is a linearized collision integral, $\vec E$ is the electric field, and $\delta f(p,r,t) = f - f_0$ describes a state weakly perturbed away from the Fermi-Dirac equilibrium state $f_0(p)=1/(e^{\beta(\epsilon(p)-\mu)}+1)$. 
Since the perturbed distribution $\delta f(p,r,t)$ is concentrated near the Fermi surface, we can expand it in the angular harmonics as 
\begin{equation}\label{eq:harmonics_delta_f_m}
	\delta f(\vec p,\vec r,t) = -\frac{\partial f_0}{\partial p}\sum_m e^{i m \theta} \delta f_m(p).
\end{equation}
where $\theta$ is the angle on the Fermi surface and the dependence on $\vec p$ in $ \delta f_m(p)$ is on the modulus of $\vec p$. 

Performing 
Fourier transform yields
\be\label{eq:delta f_m}
\delta f(\vec p,\vec r,t)=\sum_{\vec k,\omega} \delta f_{\omega,\vec k}(\vec p)e^{i\vec k\vec r-i\omega t}
,
\ee
where $\omega$ and $\vec k$ are frequencies and wavenumbers. The quantity $\delta f_{\omega,\vec k}(p)$ 
is in general a function of $\omega$ and $\vec k$ determined as discussed below. Angular decomposition of $\delta f_{\omega,\vec k}(p)$ is identical to that in Eq.\eqref{eq:harmonics_delta_f_m}:
\begin{equation}\label{eq:delta_f_m_wk}
	\delta f_{\omega,\vec k}(\vec p) = -\frac{\partial f_0}{\partial p}\sum_m e^{i m \theta} \delta f_m(p).
\end{equation}
Below we will use this expression with the angle $\theta$ on the Fermi surface measured from the direction of $\vec k$. For conciseness, we will suppress the dependence of $\delta f_m$ on $\omega$ and $\vec k$. Angular harmonics defined as in Eqs. \eqref{eq:harmonics_delta_f_m} and \eqref{eq:delta_f_m_wk} 
are eigenfunctions of a linearized collision integral,
\be 
I[e^{ i m \theta} \delta f_m(p)] = - \gamma_m \, e^{ i m \theta} \delta f_m(p)
,
\ee
where the eigenvalues $\gamma_m$ represent relaxation rates of individual harmonics of momentum distribution due to ee collisions \cite{Kryhin2022}. 

The electric field $\vec E(r)$ in Eq.\eqref{eq:BKEgeneral} can in general describe several distinct effects. It can be either applied externally, as in a calculation of electrical conduction, or be internal to the system, describing the effects of a non-equilibrium space charge distribution resulting from electron movement. 
It is therefore useful to take a moment to clarify the role of internal $\vec E$ fields in our problem. The internal fields habitually take on very different roles in 
the longitudinal and transverse response, $\vec k\parallel\vec E$ and $\vec k\perp\vec E$. In the longitudinal response, current has a finite divergence that drives density perturbations 
leading to charge piling up in the system bulk or edges and, at high frequency, excitation of collective plasma waves. For transverse response, to the contrary, current has zero divergence and density remains unperturbed. In this case, with density being constant and equal to that in equilibrium, hydrodynamic modes emerge that, at linear order, are unaffected by the internal fields. 

While here we are concerned with the transverse response which is found to be insensitive to the internal field effects, it instructive to highlight 
different roles of internal fields in the longitudinal and transverse responses. 
This 
can be done through 
a direct calculation involving a generic particle distribution perturbed away from  equilibrium and its electric field 
$\vec E(r)=-\nabla_r \int d^2r' U(\vec r-\vec r') e \delta n(\vec r')$. In Fourier representation:
\begin{equation}
\label{eq:E=-ikUn}
	\vec E(k) = -i\vec{k} U(\vec k) \sum_p\delta f_{\omega,\vec k}(p), \quad \quad U(\vec k) = \frac{2\pi e^2}{\kappa |\vec k|},
\end{equation}
where the quantity $\delta f_{\omega,\vec k}(p)$ is that in Eq.\eqref{eq:delta_f_m_wk}.
Here $\kappa$ is the dielectric constant, $\vec k$ is the spatial wavevector, and Fourier harmonics are defined in a standard manner as $\vec E(k) =\int d^2r e^{-i\vec k\vec r} \vec E(r)$. 
Eventually, we will be interested in the limit of $\omega$ and $k$ small compared to Fermi energy and Fermi wavenumber, and in the case of $\vec E$ directed perpendicular to $\vec k$. However, for the time being, we will maintain a general discussion. 
Plugging $\vec E(k)$ in Eq.\eqref{eq:BKEgeneral} and expanding particle distribution in angular harmonics, Eq.\eqref{eq:delta_f_m_wk}, yields a set of coupled equations for different harmonics $\delta f_m$. This problem describes both the longitudinal and transverse response, where $\delta f_m = \delta f_{-m}$ and $\delta f_m = - \delta f_{-m}$, respectively. 

Below we focus on the transverse case  $\vec k\perp\vec E(k)$ (and, therefore, $\delta f_m = - \delta f_{-m}$) and show that the hybridization of the $m = \pm 1$ and $m = \pm 3$ harmonics results in hydrodynamic modes with a scale-dependent 
viscosity. Since the transverse response condition $\delta f_m = - \delta f_{-m}$ implies $\delta f_0=0$, the transverse response is not accompanied by charge buildup. Accordingly, in this case the $m=0$ harmonic and the internal electric field drop out. 
These quantities, 
however, would be present for the longitudinal modes such as plasmons. Both kinds of modes, transverse and longitudinal, are readily  captured by 
writing the kinetic equation in a general form that incorporates an $\vec E$-field term: 
\begin{align}\label{eq:f_modes}
	& \left( \gamma_m - i \omega \right) \delta f_m + \frac{ikv}{2} \left( \delta f_{m-1} + \delta f_{m+1} \right) 
	\nonumber \\
	& = - \frac{ievk}{2m}\frac{\p f_0}{\p \varepsilon} U(k) \delta f_0 \left( \delta_{m,1} + \delta_{m,-1} \right),
\end{align}
where mode decay rates $\gamma_m$ obey relations $\gamma_0,\,\gamma_1 = 0$ due to particle number conservation and momentum conservation. For the moment, we carry on with the general problem, Eq.\eqref{eq:f_modes}, eventually specializing to the transverse case and discarding the $U(k)$ term. 

\section{Non-Newtonian viscosity in a two-mode model}
\label{sec:2-mode}

Here we derive hydrodynamic modes and viscosities for the two-mode model introduces above. We assume a small decay rate  for long-lived $m=3$ excitations,
\be
\gamma_3 = \gamma^\prime \ll \gamma^*={\rm min}(\gamma_5,\gamma_7,...,\gamma_{m_{\rm even}})
.
\ee 
In addition, we introduce a small momentum relaxation rate $\gamma_1=\gamma_p$ accounting for disorder and phonon scattering, assuming a small value $\gamma_p\ll\gamma^*$. 

At sufficiently long times such that $\omega \ll \gamma^*$, the even-$m$ harmonics are mostly relaxed, giving an expression that links the even harmonics and the odd harmonics:
\begin{equation}\label{eq:even_harmonics}
	\delta f_{2m} = - \frac{i k v}{2 \gamma_{2m}} \left( \delta f_{2m+1} + \delta f_{2m-1} \right).
\end{equation}
For the odd-$m$ harmonics,  
we first consider the long-lived $m=\pm 1$ and $m=\pm 3$ harmonics, which obey
\begin{align}\label{eq:1_harmonic}
	 \left( \gamma_p - i \omega \right) \delta f_{\pm1} &+ \frac{ikv}{2}\left(\delta f_0 + \delta f_{\pm2} \right) 
	\\ \nonumber
	& = - \frac{ievk}{2m}U(k)\frac{\p f_0}{\p \varepsilon} \delta f_0
\\ \label{eq:3_harmonic}
 (\gamma^\prime - i \omega) \delta f_{\pm 3} &+ \frac{ikv}{2} \left( \delta f_{\pm 2} + \delta f_{\pm 4} \right) = 0
\end{align}
The above equations are true for both the transverse and longitudinal modes. 

From now on we specialize to transverse modes. In this case, as noted above, 
density remains unperturbed, $\delta f_0 = 0$, and therefore the electric field $\vec E$ induced by density variation, Eq.\eqref{eq:E=-ikUn}, vanishes. 
Dropping $\delta f_0$ and plugging Eqs.\eqref{eq:even_harmonics} into Eq. \eqref{eq:1_harmonic}, we obtain a system of coupled equations for odd-$m$ harmonics. Retaining the harmonics $m=\pm3,\pm1$ and suppressing higher-$m$ harmonics generates coupled equations for $\delta f_{\pm 3}$ and $\delta f_{\pm 1}$ variables, 
which are valid at sufficiently long times and large distances corresponding to frequencies
\be
\omega\ll \gamma^*. 
\ee
At the same time,  $\omega$ can be either smaller or greater than the rates $\omega_p$ and $\gamma_3$, both of which are small on the scale of the higher-$m$ rates $\gamma_5$, $\gamma_7$, ..., $\gamma$. 
Introducing  notation for viscosity, $\nu = v^2/4\gamma$, 
and solving for the velocity mode $m=1$, we obtain a collective mode dispersion relation 
\begin{equation}\label{eq:omegak}
	\gamma_p - i \omega + \nu k^2 - \frac{\left( \nu k^2 \right)^2}{\gamma^\prime - i \omega + 2  \nu k^2} = 0.
\end{equation}
To clarify the behavior at long times, 
we compare Eq.\eqref{eq:omegak} to Stokes hydrodynamics $\p_t \vec v=\nu\nabla^2\vec v$.  In the limit of $\gamma_p = 0$ Eq.\eqref{eq:omegak} yields a relation for velocity mode
\begin{equation}\label{eq:scaledep}
-i\omega \delta f_1 = \Xi_{k,\omega} k^2 \delta f_1,
\end{equation}
which represents a hydrodynamic equation with a scale-dependent (non-Newtonian) viscosity
\be\label{eq:non-newtonian}
\Xi_{k,\omega}=\nu \frac{\nu k^2+\gamma^\prime - i\omega}{2\nu k^2+\gamma^\prime - i\omega}
.
\ee
The dependence on $k$ and $\omega$ in this expression indicates presence of several different regimes. 

For the regime of longest times and distances --- $|\omega|,\,\nu k^2 \ll \gamma^\prime$ --- this yields 
a viscous mode 
$i \omega =  \tilde \nu k^2$ with viscosity that has weak $k$ dependence
\be\label{eq:nu_low}
	\nu(k) = \nu \left( 1 - \frac{\nu k^2}{\gamma^\prime} +O(k^4)\right),
\ee
and one slowly-decaying mode with $i \omega \sim \gamma^\prime$.
Conversely, in the regime $\gamma^\prime \ll |\omega|,\,\nu k^2 \ll \gamma$, Eq.\eqref{eq:omegak} predicts 
two distinct viscous modes with different viscosities given in Eq.\eqref{eq:nu1,nu2}. 
As a sanity check, in the limit when the long-lived mode becomes short-lived, $\gamma^\prime \approx \gamma$, only one (the ordinary Newtonian) viscous mode survives. 

Since the number of viscous modes changes with frequency and length scale, it is instructive to consider the $k$ dependence without making any simplifying approximations. 
Eq.\eqref{eq:omegak} yields a quadratic equation for $i\omega$ that can be solved to obtain two distinct dispersing modes: 
\begin{equation}
	i \omega = \frac{1}{2} \left( \gamma_p + \gamma^\prime + 3 \nu k^2 \right)\pm \sqrt{\left( \nu k^2 \right)^2 + \frac{1}{4} \left( \gamma^\prime - \gamma_p  + \nu k^2 \right)^2}
	.
\label{eq:exactomega}
\end{equation}
This result encompasses all asymptotic regimes discussed above. Indeed, in  the long-wavelength limit $\nu k^2 \ll\gamma^\prime$, one of the modes is damped, with $i\omega$ proportional to $\gamma^\prime$ (assuming $\gamma_p \ll \gamma^\prime$). The second mode remains viscous in this limit and has a weakly dispersing viscosity described by Eq.\eqref{eq:nu_low}. At larger $k$ a pair of viscous modes with viscosities $\nu_{1,2}$ given in Eq.\eqref{eq:nu1,nu2} is recovered.


This analysis demonstrates that the long-lived excitation $\delta f_3$, coupled by $k$-linear terms to the velocity mode $\delta f_1$, generates an additional  nonclassical hydrodynamic mode. This points to the existence of a peculiar new regime positioned between the ordinary hydrodynamic and ballistic transport regimes, in which new hydrodynamic modes can emerge. These modes are expected to contribute to transport on equal footing with the conventional Stokes hydrodynamic modes. 

Lastly, a note on the validity of the analysis involving one long-lived mode. 
As discussed above, accounting for $\delta f_3$ mode and ignoring $\delta f_5$ and higher-order harmonics 
is legitimate in the interval of temperatures for which the characteristic mode frequency satisfies
\begin{equation}\label{eq:TwoModesOmegaCondition}
	\omega \ll \gamma_5
	,
\end{equation}
whereas the relation between $\omega$ and the rate $\gamma_3$ can be arbitrary. 
Since both the even-$m$ rates $\gamma_2$, $\gamma_4$, ... and the odd-$m$  rates $\gamma_7$, $\gamma_9$, ... are all greater than the rate  $\gamma_5$, the condition in Eq.\,\eqref{eq:TwoModesOmegaCondition} applies to these modes too.

\section{Hydrodynamic modes for the case of many long-lived excitations}

The disparity in even and odd excitation modes becomes significantly more pronounced as temperature $T$ decreases, as evident in Fig.\,\ref{fig:fig1} and Eq.\eqref{eq:OddAndEvenRates}. This has two important consequences. One is an increase in the number of long-lived 
odd-$m$ modes that have decay rates well under those of the even-$m$ modes with similar $m$. The number of such `active' odd-$m$ modes grows quickly as $T$ decreases. Another consequence is a rapid increase in lifetimes of these modes, described by the decay rates that scale with temperature as $\gamma\sim T^4/T_F^4$. These rates are much smaller than the typical two-body collision rates 
$\gamma_2\sim\gamma_4\sim...$ 
that define the timescale at which the hydrodynamic description emerges. 
Consequently, with a decrease in $T$, there is a rapid expansion of the family of odd-$m$ modes that contribute to hydrodynamics.
This defines an interesting phase dominated by emerging hydrodynamic modes. To gain insight into system behavior in this new regime, here we consider hydrodynamic modes in the presence of many long-lived excitations.


As a first step, we 
recall why the number of `active' long-lived 
excitations grows as $T$ decreases and estimate the number of these excitations. This happens because, as can be seen in Fig.\,\ref{fig:fig1}, odd harmonics with large values of $m$ have lifetimes comparable to those of even harmonics at a high $T$ but tend to become long-lived at a lower $T$. One can estimate the number of such long-lived harmonics by noticing that for any temperature $T$ there will be a value of $m$ that is large enough such that the corresponding odd and even harmonics are of the same order of magnitude  $\gamma_{m=2p+1} \sim \gamma_{m=2p}$. This value of $m$ sets a natural temperature-dependent limit to the number of long-lived excitations at a given temperature. In particular, the value of $m$ for which odd and even rates in Eq. \eqref{eq:OddAndEvenRates} become of the same order of magnitude is $m_* \approx \sqrt{T_F/T}$. 

In the case when several 
such long-lived excitations are present, determining the change of viscosity with scale becomes more complicated. To address this question, we consider 
an $n$-mode problem, where $n> 1$ such 
excitations 
are present simultaneously. For simplicity, at first we will take their decay rates to be negligibly small. This corresponds to frequencies in the intermediate range
\be
\gamma_p,\gamma_3...\gamma_{2n-1}<\omega<\gamma_{2n+1}
.
\ee
As above, we eliminate the fast-relaxing even-$m$ harmonics 
by substituting Eq.\eqref{eq:even_harmonics} into Eq.\eqref{eq:f_modes}. This yields a closed-form system of equations of motion for odd-$m$ excitations. For the first $n$ modes with $m=2s+1$, $s=0,1,... n-1$, these equations read
\begin{equation}\label{eq:odd_f_rates}
	- i \omega \delta f_{2s+1} = - \nu k^2 \left( 2 \delta f_{2s+1} + \delta f_{2s-1} + \delta f_{2s+3} \right)
	,
\end{equation}
provided $1\le s\le n-2$. The first and last equations, $s=0$ and $n-1$, read 
\begin{align}\label{eq:odd_f_rates_first}
- i \omega \delta f_{1} = - \nu k^2 \left( \delta f_{1} + \delta f_{3} \right)
\end{align} 
and 
\begin{align}\label{eq:odd_f_rates_last}
- i \omega \delta f_{2n-1} = - \nu k^2 \left( 2\delta f_{2n-1} + \delta f_{2n-3} \right)
,
\end{align} 
respectively. 
From the form of these equations it is evident that a total of $n$ hydrodynamic modes are present in this case, since for all solutions the $\omega$ vs. $k$ dispersion gives 
relaxation rates proportional to $\nu k^2$. Determining the spectrum of viscosities for these modes is thus equivalent to diagonalizing an $n\times n$ matrix
\begin{equation}\label{eq:matrix_nn}
	\begin{bmatrix}
		1 & 1 &  0 & \dots    & & &\\
		1 & 2 & 1 &             &  & &\\
		0 & 1 & 2 &             &  & &\\
 	\vdots  &    &    &  \ddots  &   &  &\vdots\\
		  &    &    &            & 2&  1 & 0 \\
		  &    &    &            & 1&  2 & 1 \\
		 &     &    &            \dots & 0 & 1& 2 \\
	\end{bmatrix}.
\end{equation}
The eigenstates and eigenvalues for this matrix are readily found by a comparison to a 1D tight binding problem on an infinite line
\be
H=\sum_x 2\left.|x\ra \la x|\right.+\left.|x\ra \la x+1|\right.+\left.|x+1\ra \la x|\right.
\ee
where $x$ is an integer. This problem has plane-wave eigenstates $\psi(x)=e^{ipx}$ with the eigenvalues
\[
\lambda_k=2+2\cos p
,\quad -\pi<p<\pi,
\]
degenerate in $p$ and $-p$. 

To derive the problem stated in Eq.\eqref{eq:matrix_nn}, we analyze state vectors $\psi(x)$ that fulfill two auxiliary anti-symmetry conditions specifically chosen to replicate the matrix's structure described in Eq.\eqref{eq:matrix_nn} with ``1" in the upper-left corner and ``2" in the lower-right corner.
The two anti-symmetry conditions are imposed relative to the points $x=1/2$ and $x=n+1$. Indeed, focusing on the solutions anti-symmetric relative to $x=1/2$, 
\[\psi(1)=-\psi(0),\ \psi(2)=-\psi(-1),\ ...
\] 
yields equations that are identical to those described by the upper left part of our matrix:
\[
\lambda\psi(1)=\psi(1)+\psi(2),\quad \lambda\psi(2)=2\psi(2)+\psi(1)+\psi(3),...
\]
These relations reproduce those in Eqs. \ref{eq:odd_f_rates_first} and \ref{eq:odd_f_rates}. 
At the same time, the anti-symmetry condition relative to $x=n+1$, 
\[\psi(n+1)=0,\ \psi(n)=-\psi(n+2),\ \psi(n-1)=-\psi(n+3),\  ... 
\]
yields equations identical to those described by the lower right part of the matrix:
\begin{align} \nonumber
& \lambda\psi(n)=2\psi(n)+\psi(n-1),\quad 
\\ \nonumber
& \lambda\psi(n-1)=2\psi(n-2)+\psi(n-1)+\psi(n-3),...
.
\end{align}
These relations reproduce those in Eqs. \ref{eq:odd_f_rates_last} and \ref{eq:odd_f_rates}.

Combining the two anti-symmetry conditions selects $n$ plane-wave solutions of the form $\psi(x)\sim \sin p_j(x-1/2)$ with discrete $p_j$ values
\be
p_j=\frac{2\pi j}{2n+1}, \quad j=1...n
.
\ee
%
%
The eigenvalues for these modes are $2+2\cos p_j$, yielding $n$ viscous modes with viscosity values 
\be\label{eq:viscosity_spectrum}
	\nu_j = 2 \nu \left[ 1 + \cos \left( \frac{2 \pi j}{2 n + 1} \right) \right],
	\quad
	j=1,...,n.
\ee
This result is valid for any $n\ge 2$.
As a consistency check, for $n=2$ the result found above, 
$\nu_{1,2}=\nu(3\pm\sqrt{5})/2$,  is recovered. 

These expression for $\nu_j$ imply that the largest viscosity $\nu_j$ converges to a  $n$-independent value $4\nu$ as $n$ grows. 
In contrast, the smallest viscosity $\nu_j$ scales inversely with $n^2$: $\min \nu_j \sim \nu/n^2$. Therefore, while a total of $n$ modes will be present, only a subset of those that have large enough viscosities will feature a predominantly hydrodynamic behavior. This is because for the 
low-viscosity modes the Stokes term $\nu k^2$ will have to compete with, and can be overwhelmed by, 
the non-zero decay rates $\gamma_3$, $\gamma_5$, ..., $\gamma_{2n+2}$.

We also note that a more realistic model can be constructed by letting 
different odd-$m$ harmonics to have different nonzero decay rates $\gamma_m$, and their number $n$ to scale with temperature. This generalized $n$-mode problem can be used to understand the interplay between relaxation rates for different harmonics and the number of `active' viscous modes controlling carrier transport. This interplay governs several key aspects of transport, in particular the conductance temperature dependence. 



\section{Nonlocal conductivity due to hydrodynamics modes}


Given the multitude of new hydrodynamic modes originating from long-lived excitations, it is interesting to explore how these modes impact transport in a realistic geometry.
To set the stage for this discussion, we recall that the standard treatment by Gurzhi of a viscous electron flow in a long strip predicts a temperature dependent conductance due to a viscous mode that grows with temperature as $T^2$ \cite{Gurzhi1968}. This  reflects the $T$ dependence of e-e scattering that enters the electron viscosity. The $T^2$ scaling arises when the viscosity is found without accounting for the collinear scattering effects. Collinear scattering creates long-lived excitations and results in multiple hydrodynamic modes. As demonstrated below, these modes give a unique contribution to the conductance that mimics some aspects of the conductance predicted in Gurzhi' theory 
but, in general, has a different 
dependence on system parameters such as temperature and strip width. A non-Gurzhi $T$ dependence, if observed, can provide a signature of the new hydrodynamic modes.

To address these questions, in this and the following sections we 
employ a simple model which allows for a direct analytic treatment. 
In this model, we replace the true hierarchy of even-$m$ and odd-$m$ rates $\gamma_m$ as defined in Eq.\eqref{eq:OddAndEvenRates} with a simplified hierarchy.
For several lowest harmonics, we set odd-$m$ rates to zero, whereas for the higher harmonics we  take odd-$m$ rates to be equal to even-$m$ rates. 
While this framework is not entirely realistic, it allows us to illustrate the enhancement of temperature-dependent conductivity due to multiple hydrodynamic modes. However, a more comprehensive analysis, considering the correct rate hierarchy  given in Eq.\eqref{eq:OddAndEvenRates}, reveals a temperature dependence that differs from the quadratic $T^2$ dependence found here. 
 Instead, the actual temperature dependence proves to be linear rather than quadratic, as reported in \cite{Kryhin2023b}.
 This problem, however, lies beyond the scope of this paper and will be discussed elsewhere. 

As we saw above, the long-lived excitations are manifested in additional hydrodynamic modes in an infinite system. Now we consider how these modes impact transport in a realistic geometry. Namely, we consider an infinite strip of width $w$ with diffuse carrier scattering at the boundaries and show that the extra hydrodynamic modes lead to a unique 
dependence of the DC conductivity on system parameters. To compute the electrical conductivity governed by Eq. \eqref{eq:f_modes} with the new boundary conditions we first consider an auxilary  problem of conduction in an infinite plane, 
where $\vec k$ is taken to be perpendicular to $\vec E$. 

This auxiliary problem is motivated by the following considerations: the steady-state DC current flowing in a  long strip geometry must be constant along the strip and vary 
in the direction perpendicular to the strip. Therefore, in this geometry all current-carrying modes have $\vec k$ perpendicular to the strip and $\vec E$ along the strip.  To described conduction in this system, we will add an extra term to Eq.\eqref{eq:f_modes} that describes a fictitious internal electric field localized at the strip boundary which couples to $\delta f_m$ modes. 
In general, the field $\vec E$ 
enters transport equations through two
terms on the right-hand side,  
\begin{equation*}
	- \frac{e v_F}{2} \frac{\p f_0}{\p \varepsilon} \left[ E_\parallel \left(\delta_{m,1} + \delta_{m,-1}\right) + i E_\perp \left(\delta_{n,-1} - \delta_{n,+1}\right) \right],
\end{equation*}
where $E_\parallel$ and  $E_\perp$ are the components of $\vec E$ along 
and perpendicular to $\vec k$.
However, 
since in our case $\vec E$ is perpendicular to $\vec k$, we can set $E_\parallel = 0$ and $E_\perp = E_0$. 

Combined with the transverse mode condition $\delta f_m = - \delta f_{-m}$, equations for $\delta f_m$, Eq.\eqref{eq:f_modes}, can be solved recursively [see Appendix and Ref.\cite{Nazaryan2021}], yielding a continued fraction expression for conductivity 
\begin{equation}\label{eq:sigma_transverse}
	\sigma_\perp(k) = \frac{n_0e^2}{m}
		\frac{1}{\gamma_p + \frac{z}{\gamma_2 + \frac{z}{\gamma_3 +\frac{z}{\gamma_4 + ...}}} }
		,\quad
		z=v_F^2 k^2/4
		,
\end{equation}
where we expressed carrier density $n_0=g p_F^2/4 \pi \hbar^2$ through Fermi momentum and set $\omega = 0$ for the DC response (here $g$ is the spin/valley degeneracy). Starting with this general result, which is true for any decay rate values $\gamma_m$, allows to incorporate our analysis in a wider context. The $k$-dependent conductivity describes a nonlocal current-field conductivity response with the nonlocality accounting for carrier movement accompanied by velocity distribution relaxation with the rates $\gamma_m$ that are different for different angular harmonics. 

The result in Eq.\eqref{eq:sigma_transverse} can be readily applied to the 
$n$-mode problem with 
$n$ long-lived excitations introduced above, for which the viscous mode spectrum, Eq.\eqref{eq:viscosity_spectrum}, was derived. In this model the odd-$m$ rates $\gamma_3$, $\gamma_5$, ..., $\gamma_{2n+1}$ were taken to be zero, whereas for all other modes the rates equal $\gamma$ except for the $m=1$ rate that takes a small nonzero value $\gamma_1=\gamma_p\ll\gamma$. 
For $i>2n+1$, the tail of the 
continued fraction in Eq.\eqref{eq:sigma_transverse}, i.e. the infinite part of the continued fraction starting with $\gamma_{2n+1}$, can be evaluated exactly with the help of the identity
\begin{equation}\label{eq:fraction_tail}
	\frac{z}{\gamma + \frac{z}{\gamma + \frac{z}{\gamma + ...}}} = \frac{\gamma}{2} \left( \sqrt{1 + \frac{4z}{\gamma^2}} - 1\right). 
\end{equation}
Plugging this result in Eq.\eqref{eq:sigma_transverse} gives
\begin{align}\label{eq:sigma_transverse_simplified}
	\sigma_\perp(k) & = \frac{n_0e^2}{m}
		\frac{1}{\gamma_p + \frac{z}{(n+1)\gamma + \frac{\gamma}{2} \left( \sqrt{1 + \frac{4z}{\gamma^2}} - 1\right)
		} }
		\\
		& \approx
		\frac{n_0e^2}{m}
		\frac{1}{\gamma_p + \frac{z}{(n+1)\gamma 
		} }
\end{align}
In the last line we 
suppressed the quantity evaluated in Eq.\eqref{eq:fraction_tail}, 
which is small in the regime of interest 
\be\label{eq:condition_1}
v_F k \ll n \gamma
\ee
that is $k$ values that lie outside the ballistic transport regime. 
For a realistic estimate, we reintroduce nonzero values of odd-$m$ rates $\gamma_3$, ..., $\gamma_{2n+1}$, which were previously taken to be zero. 
Taking these values to be small, $\gamma^\prime \ll \gamma$, we arrive at the multimode hydrodynamic regime occurring at the lengthscales 
\be\label{eq:condition_2}
{\rm max}(\gamma^\prime, n \gamma_p) \ll \nu k^2 \ll n^2 \gamma
\ee 
in between the conventional hydrodynamic and ballistic regimes. At these lengthscales the effects of additional viscous modes become essential. 
The dependence on the number of long-lived modes $n$ in Eq.\eqref{eq:condition_2} arises as follows. The factor $n^2$ is due to the condition for ballistic transport, Eq.\eqref{eq:condition_2}. The factor $n$ multiplying $\gamma_p$ is read off Eq.\eqref{eq:sigma_transverse_simplified}, taking the limit $n\gg 1$. 
The dependence on $\gamma^\prime$ in Eq.\eqref{eq:condition_2} can be inferred from the analysis of the fraction 
in Eq. \eqref{eq:sigma_transverse}. In deriving the phase boundary for the crossover to single-mode hydrodynamics (the  inequality $\gamma'<\nu k^2$) we assumed identical decay rates equal $\gamma'$ for all long-lived modes. 

Setting $\gamma_p=0$ in Eq.\eqref{eq:sigma_transverse_simplified} yields a simple
expression for $k$-dependent conductivity: 
\begin{equation}\label{eq:sigma_perp_final}
	\sigma_\perp(k) = \frac{e^2 n_0}{m} \frac{n+1}{\nu k^2}. 
\end{equation}
This result coincides in form with the conductivity derived from Stokes equation multiplied by a factor $n+1$, reflecting an enhancement in conductivity due to the presence of the additional 
hydrodynamic modes. Interestingly, while the viscous modes found above feature a wide distribution of viscosity values, Eq.\eqref{eq:viscosity_spectrum}, their net contribution to conduction carries little information about this complexity. Instead, it matches the conduction of a single conventional hydrodynamic mode, recovered from Eq.\eqref{eq:viscosity_spectrum} at $n=0$, multiplied by the number of the long-lived modes $n+1$. 


\section{Transport in a strip geometry}

Next, we use the expression for $\sigma_\perp(k)$ found in an infinite plane, Eq.\ref{eq:sigma_perp_final}, 
to determine current distribution and conductivity in a long strip, 
\[
0<x<w,\quad -\infty<y<\infty ,
\]
where the axis $x$ is directed perpendicular to the strip, whereas the electric field and current are directed along the strip. To tackle the boundary conditions we employ the method used in Ref.\cite{tomogrph}, wherein  
a fictitious electric field is introduced at the strip boundaries $x=0$, $w$ of value adjusted 
to achieve zero current at the boundary. Physically, this electric field 
describes dissipation due to carrier momentum relaxation at the boundary. Writing this auxiliary field as $\delta E=-\alpha j$, with $j$ a yet unknown current at the boundary and the minus sign describing momentum relaxation, 
yields a self-consistent current-field relation for the total field
\be
E'(x) = E_0 +\delta E
\ee
which includes the contribution of current dissipated at the boundary. It is convenient to extend the strip geometry periodically to the entire plane, writing the relation between the field $E'(x)$ and current $j(x)$ as
\begin{equation}\label{eq:Eperp}
	E'(x) = E_0 - \alpha \sum_n j(x_n) \delta(x - x_n)
	. 
\end{equation}
Here $x_n = wn$ describes strip boundaries periodically replicated throughout $-\infty<x<\infty$. 
The electric field and current, directed along the strip, vary across the strip but are constant along the strip. The advantage of such infinite-space setting is that the relation between current and field is 
provided by the quantity $\sigma_\perp(k)$ discussed above.

The idea of the method is to solve for the self-consistent response of the system of the current $j(x)$ to the field $E'$ and express the result only as a function of $E_0$.
To that end we assume a current-dependent $E'$ that differs from an externally applied $E$ by a boundary term as given above, and solve Eq.\eqref{eq:Eperp} together with a nonlocal current-field relation
\be\label{eq:integral_equation}
j(x)=\int_{-\infty}^\infty dx' \sigma(x-x') E'(x')
.
\ee
Here $j$ and $E'$ are the current and field $y$ components and we introduced a  nonlocal conductivity kernel $\sigma(x-x')=\int_{-\infty}^\infty \frac{dk}{2\pi}e^{ik(x-x')}\sigma_\perp(k)$. The mathematical procedure involves solving for the current response governed by nonlocal conductivity 
for $j(x)$ as a function of $\alpha$ and $E_0$. Subsequently, taking the limit of $\alpha \rightarrow \infty$ ensures that the current vanishes at the boundaries $x=x_n$. This yields a one-dimensional integral equation, Eq.\eqref{eq:integral_equation}, for the current distribution on a line $-\infty<x<\infty$, which is periodic in $x$ with periodicity $w$.
The solution of this integral equation describes current distribution that is governed by $\sigma_\perp(k)$ within the strip and, in the limit $\alpha \rightarrow \infty$, vanishes at the boundary. 

In the infinite strip problem, both the current and the electric field, which are oriented along the strip, vary in the direction perpendicular to the strip. Consequently, there is no space charge buildup due to current flow in the system, allowing us to disregard longitudinal conductivity and work with the transverse conductivity $\sigma_\perp(k)$.

The integral equation, Eq.\eqref{eq:integral_equation}, can be solved for any $k$ dependence $\sigma_\perp(k)$ by Fourier transform [see Ref.\cite{tomogrph}]. 
Since our expression for $\sigma_\perp(k)$ in Eq. \eqref{eq:sigma_perp_final} has the same $1/k^2$ form as the viscous conductivity in Stokes 
hydrodynamics, the results of Ref.\cite{tomogrph} can be applied directly. 
This yields a current distribution $j(x)$ within the strip that has a parabolic profile familiar for viscous Poiseuille flow. 
However, because of a simultaneous presence of $n$ hydrodynamic modes the current-field relation is distinct from Gurzhi's 
hydrodynamic-limit conductivity: 
\begin{equation}\label{eq:Gurzhi_times_(n+1)}
	j(x) = \sigma_\mathrm{eff} E_0 \frac{x}{w} \left( 1 - \frac{x}{w} \right), \quad \sigma_\mathrm{eff} = \frac{\gamma e^2 m w^2}{12 \pi \hbar^2}(n+1).
\end{equation}
This result coincides with the Gurzhi's 
conductivity 
enhanced by a factor $n+1$, whereas in the absence of the odd-$m$ long-lived modes, $n = 0$, it matches exactly the result of Gurzhi's theory. 
 
Below we use these results to describe different regimes arising in our problem. We first consider temperature $T$ fixed and strip width $w$ varying, for simplicity keeping $\gamma_p=0$. Since for a flow in a strip the characteristic wavenumber is $k \sim 1/w$, 
the validity condition for the result in Eq.\eqref{eq:Gurzhi_times_(n+1)} follows directly from $\gamma^\prime \ll \nu k^2 \ll n^2 \gamma$, giving
\begin{equation}\label{eq:condition_3}
	\sqrt{\gamma^\prime \gamma} \ll \frac{v_F}{w} \ll n \gamma.
\end{equation}
These inequalities define three distinct transport regimes 
that occur for different strip width values. Namely:
\begin{enumerate}[label=\arabic*)]
\item
Ballistic regime occurs for relatively small strip widths 
$w$ described by the breakdown of the right inequality in Eq.\eqref{eq:condition_3}, 
\begin{equation}
	n \gamma \ll \frac{v_F}{w}.
\end{equation}
In this regime, the conductivity of the strip is large. 
Because of the factor of $n$, the range of $w$ where ballistic regime can occur is reduced. Since $\gamma\sim T^2$, at a fixed $w$ and upon varying $T$ this regime is pushed into lower temperatures as compared to the conventional single-mode hydrodynamics. 
\item
The multi-mode hydrodynamic regime occurs for $w$ satisfying the condition in Eq.\eqref{eq:condition_3}. 
In this regime the conductivity 
will be enhanced by a factor of $(n+1)$ in comparison to the conventional hydrodynamic regime:
\begin{equation}
	\sigma_\mathrm{eff} = \frac{\gamma e^2 n_0 w^2}{3 m v_F^2}(n+1).
\end{equation}
\item
The conventional hydrodynamic regime occurs when $w$ values are very large, such that
\begin{equation}
	\frac{v_F}{w} \ll \sqrt{\gamma^\prime \gamma}.
\end{equation}
Conductivity in this regime is described by the standard Gurzhi result:
\begin{equation}\label{eq:sigma_eff_n}
	\sigma_\mathrm{eff} = \frac{\gamma e^2 n_0 w^2}{3 m v_F^2}.
\end{equation}
\end{enumerate}
We therefore see that 
as $w$ grows the system does not transition directly from a ballistic into a hydrodynamic regime. Instead, it features a new regime of an enhanced conductivity. 
Understanding the conductivity temperature dependence in this regime for a realistic system is an interesting problem which requires 
accounting for the fact 
that both the decay rates $\gamma$ and $\gamma^\prime$ as well as 
the number of long-lived modes $n$ depend on temperature. As discussed in Ref.\cite{Kryhin2023b}, this opens a possibility of a new temperature scaling of $\sigma_\mathrm{eff}$ distinct from the $T^2$ scaling predicted from Eq.\eqref{eq:sigma_eff_n}.
 
However, in the model with a $T$-independent $n$ considered here,  the temperature scaling remains 
of the same 
form as in the Gurzhi regime, $\sigma_\mathrm{eff}\sim T^2$. The only difference in conductivity is  
an $(n+1)$-fold increase relative to the Gurzhi conductivity. 
Therefore, one would only find non-trivial temperature scaling at the parameter values that correspond to the transition between different regimes. 

Another interesting piece of information provided by this model are the temperature-dependent boundaries between different regimes.
Taking the temperature dependence of the rates $\gamma$ and $\gamma^\prime$ to be
\begin{equation}
	\gamma = g \frac{T^2}{T_F^2}, \quad \gamma^\prime = g \frac{T^4}{T_F^4}
	,
\end{equation}
we predict that the multi-mode hydrodynamic regime occurs for $T$ values satisfying
%
\begin{equation}
	g \frac{T^3}{T_F^3} \ll \frac{v_F}{w} \ll g n \frac{T^2}{T_F^2}.
\end{equation}
Therefore, the ballistic regime will occur for $T \ll T_F \sqrt{v_F/g w n^2}$, whereas the 
multimode hydrodynamics will occur at intermediate temperatures 
\be
T_F \sqrt{v_F/g w n^2} \ll T \ll T_F \sqrt[3]{v_F/g w}
.\ee
This temperature interval can be wide provided $v_F/g w \ll 1$. Finally, at the largest temperatures $T_F \sqrt[3]{v_F/g w} \ll T$ the conventional Gurzhi hydrodynamics will occur. As above, 
these estimates intentionally ignore momentum relaxation processes for the sake of simplicity. Accounting for a finite value of $\gamma_p>0$ can be straightforwardly achieved by confining the analysis to length scales over which the hydrodynamic modes propagate during timescales of the order of $t\sim \gamma_p$.

A more crucial factor overlooked by the above analysis is temperature dependence of the number $n$ of `hydrodynamically active' modes originating from long-lived excitations in a realistic system. As illustrated in the hierarchy of decay rates shown in Fig.\,\ref{fig:fig1}, it is entirely reasonable to anticipate a rapid increase in the number of these modes as the temperature $T$ decreases. A comprehensive analysis of these effects \cite{Kryhin2023b} predicts that the conductivity scales linearly with temperature, $\sigma\sim T$. This linear scaling persists down to the lowest temperatures. This behavior can be reconciled with the above estimates provided $n$ varies inversely with $T$, obeying an inverse relationship $n\sim 1/T$. Notably, as emphasized in Ref.\cite{Kryhin2023b}, this prediction finds support in experimental observations. The linear-in-$T$ scaling of the hydrodynamic conductivity is  a striking prediction of multimode hydrodynamics that stands in sharp contrast with the conventional electron hydrodynamics.

In conclusion, the quasiparticle scattering in 2D Fermi gases exhibits a highly collinear character due to fermion exclusion and kinematic constraints, even when the angle dependence of microscopic interactions is negligible. This unique kinetic behavior has relevance across various 2D systems, particularly in cases where electron-electron collisions represent the dominant scattering mechanism, overshadowing other carrier relaxation pathways. As a result, the carrier dynamics in these systems give rise to long-lived excitations, which, in turn, lead to novel hydrodynamic modes characterized by non-Newtonian (scale-dependent) viscosity. 
This leads to emergence of multiple viscous modes that are not encountered in conventional fluids. 
These multiple viscous modes create new transport regimes that offer clear and testable signatures of the distinctive behavior stemming from long-lived excitations at a 2D Fermi surface.

This work was supported by the Science and Technology Center for Integrated Quantum Materials, National Science Foundation grant No.\,DMR1231319 
and was performed in part at Aspen Center for Physics, which is supported by 
NSF grant PHY-2210452. 
SK is currently affiliated with Harvard University, Physics Department.


\begin{appendix}

%
%
%
%

\section*{\label{sec:sigma}Appendix: Scale dependent conductivity and continued fractions}
\setcounter{equation}{0}
Here we derive a general relation between nonlocal conductivity $\sigma(k)$ and the relaxation rates $\gamma_m$ for the  angular harmonics of carrier distribution, which is 
used in Eq.\ref{eq:sigma_transverse} and subsequent discussion in the main text. 
While the general form of a continued-fraction representation of $\sigma(k)$ is identical to that obtained elsewhere \cite{Nazaryan2021}, it is summarized here for reader's convenience. 

As a starting point, we use the Boltzmann kinetic equation for electrons in the presence of an external electric field, linearized in small deviations of carrier distribution from equilibrium: 
\begin{align}\refstepcounter{equation}\tag{A\arabic{equation}}
 & \left(\partial_{t}+\vec{v}\nabla_{\vec{x}}-I\right)\delta f_{\vec{p}}(t,\vec{x})=-e\vec{E}\nabla_{\vec{p}}f_{\vec{p}}^{(0)}
 ,
\end{align}
where 
$I$ is the collision operator, $f_{\vec{p}}^{(0)}$ is the equilibrium distribution, and the electric field term can be expressed through carrier velocity as 
$\vec{E}\nabla_{\vec{p}}f_{\vec{p}}^{(0)}=\vec{E}\vec{v}\frac{\partial f_{\vec{p}}^{(0)}}{\partial\epsilon}$. 
The perturbed distribution can be decomposed into a sum of cylindrical
harmonics as $\delta f_{\vec{p}}=e^{i\vec{k}\vec{x}-i\omega t}\sum_{m}\delta f_{m}e^{im\theta}$, where $\theta$ is the azimuthal angle on the Fermi surface. Due to the cylindrical symmetry, the harmonics $\delta f_{m}e^{im\theta}$ are eigenfunctions of the linearized collision operator,
\be \refstepcounter{equation}\tag{A\arabic{equation}}
I \delta f_{m}e^{im\theta}=-\gamma_m \delta f_{m}e^{im\theta}
,
\ee 
where $\gamma_{m}$ are relaxation rates originating from microscopic processes of carrier scattering and collisions. In particular, $\gamma_1=\gamma_{p}$ describes momentum relaxation due to disorder of phonon scattering, $\gamma_2$ is due to electron-electron collisions, $\gamma_0=0$ due to particle number conservation, and so on.  
The key observation is that using the basis $\delta f_{m}e^{im\theta}$ the problem can be brought to the form described by a tridiagonal matrix, a representation in which a closed-form solution for conductivity $\sigma(k)$ can be given in terms of continued fractions. This representation is obtained by noting that the terms $\vec{v} \vec{k}$ and $e \vec{v} \vec{E}$, when rewritten in the angular harmonics basis, have nonzero matrix elements only between harmonics $m$ and $m\pm 1$. This is made apparent by the identities 
\begin{align}
\vec{v} \vec{k}&=\frac{v}{2}\left(k_{x}+i k_{y}\right) e^{-i \theta}+\frac{v}{2}\left(k_{x}-i k_{y}\right) e^{i \theta}
\nonumber \\
&=\zeta e^{-i \theta}+\bar{\zeta} e^{i \theta} \refstepcounter{equation}\tag{A\arabic{equation}}\\
e \vec{v} \vec{E}&=\frac{ev}{2}\left(E_{x}+i E_{y}\right) e^{-i \theta}+\frac{ev}{2}\left(E_{x}-i E_{y}\right) e^{i \theta}
\nonumber \\
&=\mathcal{E} e^{-i \theta}+\bar{\mathcal{E}}e^{i \theta}\refstepcounter{equation}\tag{A\arabic{equation}}
,
\end{align}
where we introduced notation $\zeta=v(k_x+ik_y)/2$ and $\mathcal{E}=ev\left( E_x+i E_y\right)/2$. 
Accordingly, the Boltzmann equation turns into a system of coupled linear equations:
\begin{align}
\gamma_m \delta f_m+ \zeta\delta f_{m+1}+\bar{\zeta} \delta f_{m-1}= \frac{\partial f_\textbf{p}^{(0)}}{\partial \epsilon} \left(\mathcal{E} \delta_{m,-1} + \bar{\mathcal{E}}\delta_{m,1}\right) \refstepcounter{equation}\tag{A\arabic{equation}}
.
\end{align}
This problem describes a response of variables $\delta f_m$ to the ``source'' $\mathcal{E} \delta_{m,-1} + \bar{\mathcal{E}}\delta_{m,1}$. 

To solve these equations, we first limit the discussion to electric field perpendicular to the wave vector. We consider the source term with $m=1$, adding the contribution of the source term with $m=-1$ later. We introduce $\alpha_m=i \delta f_{m+1}/\delta f_m$, which brings equations with $m>1$ to the form
\begin{align} \refstepcounter{equation}\tag{A\arabic{equation}}
  \gamma_m+\zeta\alpha_m-\frac{\bar{\zeta}}{\alpha_{m-1}}=0
  .
  \end{align}  
  This equation can be rewritten as a recursion relation $\alpha_{m-1}=\frac{\bar{\zeta}}{\gamma_m+\zeta\alpha_m}$ which can be solved iteratively over $m+1$, $m+2$,$\dots$ yielding a continued fraction
  \begin{align} \refstepcounter{equation}\tag{A\arabic{equation}}
  \alpha_{m-1}=\frac{\bar{\zeta}}{\gamma_m+\frac{\left|\zeta\right|^2}{\gamma_{m+1}+\frac{\left|\zeta\right|^2}{\gamma_{m+2}+\dots}}}
  .
  \end{align}
  Similarly, for $m<1$ we define $\alpha'_m=i \delta f_{m-1} / \delta f_{m}$ and obtain 
  \begin{align} \refstepcounter{equation}\tag{A\arabic{equation}}
  \alpha'_{m+1}=\frac{\zeta}{\gamma_{m}+\frac{\left|\zeta\right|^{2}}{\gamma_{m-1}+\frac{\left|\zeta\right|^{2}}{\gamma_{m-2}+\ldots}}}
  .
  \end{align}
  Now, the harmonic $\delta f_{1}$ can be found from the $m=1$ equation 
  \be \refstepcounter{equation}\tag{A\arabic{equation}}
  \gamma_{1} \delta f_{1}+i \zeta \delta f_{2}+i \bar{\zeta} \delta f_{0}=\frac{\partial f_\textbf{p}^{(0)}}{\partial \epsilon}{\mathcal{E}}.
  \ee 
  Rewriting it as $\delta f_{1}\left(\gamma_{1}+\zeta \alpha_{1}+\bar{\zeta} \alpha'_{1}\right)=\frac{\partial f_\textbf{p}^{(0)}}{\partial \epsilon}\mathcal{E}$ and substituting the continued fractions for $\alpha_{1}$ and $\alpha'_{1}$ yields 
\begin{align}\label{eq:f1_contfrac}
\delta f_{1}&=\frac{\partial f_\textbf{p}^{(0)}}{\partial \epsilon}\frac{\mathcal{E}}{\gamma_1+\frac{|\zeta|^{2}}{\gamma_{2}+\frac{|\zeta|^{2}}{\gamma_{3}+\frac{|\zeta|^{2}}{\gamma_{4}++\ldots}}}+\frac{|\zeta|^{2}}{\gamma_{0}+\frac{|\zeta|^{2}}{\gamma_{-1}+\frac{|\zeta|^{2}}{\gamma_{-2}+\ldots}}}}
\nonumber\\
&=\frac{\partial f_\textbf{p}^{(0)}}{\partial \epsilon}\frac{\mathcal{E}}{2\gamma_1+\frac{2|\zeta|^{2}}{\gamma_{2}+\frac{|\zeta|^{2}}{\gamma_{3}+\frac{|\zeta|^{2}}{\gamma_{4}+\ldots}}}}
, \refstepcounter{equation}\tag{A\arabic{equation}}
\end{align}
where we used the identities $\gamma_{-m}=\gamma_m$ and $\gamma_{0}=0$,  accounting for the inversion symmetry and particle number conservation. The $m=-1$ source term contribution, found in a similar manner, is given by an expression identical to Eq.\eqref{eq:f1_contfrac} in which $\mathcal{E}$ is replaced with $\bar{\mathcal{E}}$.

Now it is straightforward to obtain the nonlocal conductivity by combining the current density $j_{y,k}=e v \nu_{0} \oint(d \theta / 2 \pi) \sin \theta \delta f(\theta)$ and the definition of conductivity $\vec{j}_{\vec{k}}=\sigma(k, \omega) \vec{E}_{\vec{k}}$. In the static limit $\omega=0$ we find 
\begin{align} \refstepcounter{equation}\tag{A\arabic{equation}}
& \sigma(k)=\frac{D}{\gamma_{1}+\Gamma(k)},\quad \Gamma(k)=\frac{z}{\gamma_2+\frac{z}{\gamma_3+\frac{z}{\gamma_4+\ldots}}}
,
\end{align}
with $ D=ne^{2}/m$ the Drude weight and $z=k^{2}v^{2}/4$. 

It is instructive to consider an example of equal rates $\gamma_2=\gamma_3=\gamma_4=
...=\gamma$ for which 
the quantity $\Gamma(k)$ can be evaluated in a closed form. 
Using a recursion relation $\Gamma(k)=z/(\gamma+\Gamma(k))$ we find
\begin{align} \refstepcounter{equation}\tag{A\arabic{equation}}
 & \Gamma(k)=\frac{-\gamma+\sqrt{\gamma^{2}+k^{2}v^{2}}}{2}
 .
\end{align}
This gives a scale-dependent conductivity 
\begin{align} \refstepcounter{equation}\tag{A\arabic{equation}}
 & \sigma(k)=\frac{2D}{2\gamma_{p}+\sqrt{v^2 k^{2}+\gamma^2}-\gamma}
 \label{sigma}
,
\end{align}
where we replaced $\gamma_1$ with the momentum relaxation rate $\gamma_{p}$ to make our notations agree with those in the main text. The model describes the viscous and ballistic regimes, and the crossover between these regimes at the lengthscales corresponding to $kv\sim\gamma$.

So far we considered the transverse conductivity arising for $\vec E\perp \vec k$. A straightforward generalization of the above derivation for an arbitrary orientation of electric field relative to the wave vector yields an additional tensor structure $\delta_{\alpha \alpha^\prime} - \hat{k}_\alpha \hat{k}_{\alpha^\prime}$.
This tensor structure accounts for the fact that the electric field component parallel to the wave vector $\vec k$ is screened out and does not produce a DC current. This tensor structure is irrelevant for the strip geometry considered in the main text, since in this case the field and current are directed perpendicular to the characteristic wavevector, it can be important in other geometries of interest, such as transport in a constriction or a Corbino geometry. 
\end{appendix}
\end{document}